\documentclass[runningheads]{llncs}
\usepackage[utf8]{inputenc}
\usepackage{longtable}
\usepackage{supertabular,booktabs}
\usepackage{caption}
\usepackage{multirow}
\usepackage{graphicx}
\usepackage{float}
\usepackage{comment}

\begin{document}

\title{Remote Working Pre- and Post-COVID-19: \\An Analysis of New Threats and Risks to Security and Privacy \thanks{This is the author version of the article: Nurse J.R.C., Williams N., Collins E., Panteli N., Blythe J., Koppelman B. (2021) Remote Working Pre- and Post-COVID-19: An Analysis of New Threats and Risks to Security and Privacy. In: Stephanidis C., Antona M., Ntoa S. (eds) HCI International 2021. HCII 2021. Communications in Computer and Information Science, vol 1421. Springer, Cham. https://doi.org/10.1007/978-3-030-78645-8{\_}74}}

\titlerunning{Remote Working Security and Privacy Risks}

\author{
Jason R.C. Nurse\inst{1} \and
Nikki Williams\inst{2} \and
Emily Collins\inst{3} \and
Niki Panteli\inst{4} \and
\\John Blythe\inst{5}  \and
Ben Koppelman\inst{6} 
}
\authorrunning{Nurse et al.}

\institute{University of Kent, UK \\
\email{J.R.C.Nurse@kent.ac.uk}
\and
Cranfield University, UK
\and
Cardiff University, UK
\and 
Royal Holloway, University of London, UK
\and
CybSafe, London, UK
\and
CyberSmart, London, UK\\
}

\maketitle              
\begin{abstract}
COVID-19 has radically changed society as we know it. To reduce the spread of the virus, millions across the globe have been forced to work remotely, often in make-shift home offices, and using a plethora of new, unfamiliar digital technologies. In this article, we critically analyse cyber security and privacy concerns arising due to remote working during the coronavirus pandemic. Through our work, we discover a series of security risks emerging because of the realities of this period. For instance, lack of remote-working security training, heightened stress and anxiety, rushed technology deployment, and the presence of untrusted individuals in a remote-working environment (e.g., in flatshares), can result in new cyber-risk. Simultaneously, we find that as organisations look to manage these and other risks posed by their remote workforces, employee's privacy (including personal information and activities) is often compromised. This is apparent in the significant adoption of remote workplace monitoring, management and surveillance technologies. Such technologies raise several privacy and ethical questions, and further highlight the tension between security and privacy going forward. 

\end{abstract}

\section{Introduction}

The impact that COVID-19 has had on society is undeniable. Countries, companies and individuals have had to drastically change the way they engage and operate to preserve life and prioritise safety~\cite{whoimp2021}. Technology-enabled remote working, in particular, has seen a substantial increase with millions across the globe being forced to work from at home. While remote working is not novel, the extent and speed at which it has been implemented over the last year is noteworthy as there are several compounded HCI, security and privacy implications for individuals and their employers. For instance, new forms of cybercrime and misinformation have emerged during the pandemic---both at the point of the initial spread (with online fraud and scams exploiting increased anxiety and poor mental health), and now as vaccines are being administered (with online anti-vax campaigns)~\cite{LALLIE2021102248,naidoo2020multi}.  

In this paper, we report on a critical analysis of the technology-related security and privacy issues arising due to the large-scale move to remote working as a result of COVID-19. This is based on openly available reports, media and academic articles. We consider risks that have arisen due to the increase in make-shift offices at home, the extent of distractions accompanying remote-working environments, and the abrupt adoption of various forms of new technology/apps to interact (e.g., Zoom, Microsoft Teams, Clubhouse, Houseparty). Of particular interest is how cyber security concerns and solutions have shifted before and after COVID-19. This incorporates the upsurge in attacks targeting remote work forces and challenges companies have had securing remote workforces (some of which pertain to difficulties in human use of new technology)~\cite{buil2021cybercrime,infosecmag2020}. Finally, we discuss security implications for the future as remote working and the technologies that support it are likely to become further embedded into society, including workplaces, education institutions and business.

\section{Research Methodology} 
\label{sec:Methodology}

The methodology that we adopt for this research is based on a critical review of current literature, particularly reports from industry over the last year. To direct our study into the actual security and privacy risks to remote working pre- and post-COVID-19, we conduct an online search for current reports and articles around three core areas. The first area seeks to consider the characteristics and features of work-from-home scenarios, and the impact of COVID-19 on employees. The second area examines the security risks that emerge due to characteristics and features of remote working scenarios and COVID-19. The last area pays attention to the privacy risks emerging and the tensions with organisations aiming to secure remote systems, and employees seeking to maintain some privacy in their home environments. Once these articles were identified, they were then assessed to extract key issues, especially ones that have surfaced due to the pandemic. The sections below report the key findings and results from these analyses.

\section{Remote Working since COVID-19} 
\label{sec:Workingfromhome}

Remote working, and working from home, have become the norm for many due to the pandemic. Such working has been required by governments due to various national and regional lockdowns, and many companies have encouraged this practice even after governmental mandates. Remote working is, of course, not new and has existed for a long time due to its many advantages including flexible working, increased time with family, and better work-life balance~\cite{wef2020wfh}. From our research we found that many of these advantages were upheld during the pandemic. There were also some new additions such as an increased feeling of safety, due to lack of the need to commute to work or enter an office or public space, which may put one's health at risk. 

The reported concerns accompanying remote working vastly outweighed the benefits in literature. The most commonly appearing were distractions from home life (e.g., family members, pets, chores) and friends, feelings of isolation, difficulties in communicating and team working with colleagues, overworking due to the desire to prove that one is working, technology problems, lack of visibility of staff, and  difficulty finding an appropriate work-life balance~\cite{green2020working,insider2019work,tremblay2012telework}; all of which can impact productivity. 

Assessing the influence of COVID-19 on remote working, we found that the pandemic exacerbated many of the existing challenges with this type of work. Isolation, burnout, and difficulties managing and supporting remote teams were high on the list of organisational and employee issues. There were some key differences in remote working due to COVID-19 however. For instance, COVID-19 forced employees to work from home, instead of it being a voluntary decision; this led to a completely new working experience for millions who had never worked remotely before~\cite{ibm2020wfhs,waizenegger2020affordance}. Also, employers and employees had little time to prepare for the mass need for remote working~\cite{waizenegger2020affordance}. This meant that technology facilities (e.g., laptops, home offices or teleworking software) were often not in place, some technologies had to be rapidly adopted without proper testing (therefore increasing demands on technical support staff as well), and that other important concerns such as family commitments (e.g., new childcare or elderly-care demands) and well-being (both mental and physical) were neglected. These issues were particularly salient given the overall increased negative impact on mental health, job security and finances due to the pandemic~\cite{wilson2020job}. 

\section{Security Risks} 
\label{sec:SecurityRisks}

Cyber security has been a key concern during the pandemic as companies have been rushed into migrating to new technology platforms and services to communicate, allow remote working (and remote access to corporate systems) and for business engagement. Cyber criminals have kept track of the various issues caused by remote working, as well as the general pandemic, to increase their variety and number of attacks~\cite{LALLIE2021102248}. We examined the challenges to remote working in the context of security vulnerabilities and threats to identify a set of noteworthy risks emerging specifically due to, or greatly exacerbated by, the COVID-19 pandemic. These are arranged into two main areas, security risks associated with employees working remotely, and those related to the technologies that have been in use during the pandemic. 

Employee-related security risks are focused on those issues that may target or be caused  (intentionally or unintentionally) by an employee. We list exemplars of the key risks below.

\begin{itemize}
    \item \textit{Increased likelihood of falling victim to cyber-attacks (e.g., phishing) because of a lack of concentration or distractions caused by a home-working space}. This may link to family responsibilities or household needs that are new because of the pandemic (e.g., home schooling, entire families or flatmates at home for extended periods). 
    \item\textit{Lack of remote-working security training resulting in poor security practices} that increase the potential of a compromising cyber-attack. Many organisations were not able to train employees adequately before they were forced to work from home, which compounds this risk. 
    \item \textit{De-prioritisation of security as a key concern because of heightened anxiety, stress, depression, burnout and poor mental health generally motivated by the pandemic}. As individual employees focus more on basic needs (e.g., safety, health, job security), they may be less cognisant of workplace security concerns.
    \item \textit{Reduced access to information/knowledge that causes poor security practices}, for instance, difficulty in quickly speaking with a work colleague about appropriate security behaviours when faced with a security-related decision. This issue is exacerbated by the length of time at which employees have had to work remotely, and the psychological differences of `popping by' a colleague's desk or `disrupting their work' by requesting a video call. 
    \item \textit{Trusted/untrusted individuals in the remote-working environment (or household)} may exploit new access to corporate data or services (e.g., using an unlocked laptop or phone, or listening to a confidential phone call). The reality is that these environments may be shared with unknown flatmates or others that may use this extended home-working period for malevolent purposes.
    \item \textit{Employees now experiencing minimal management monitoring or oversight may use that opportunity to steal confidential information from their employer or misuse corporate services}. This may be further motivated by perceived job insecurity due to the pandemic; a period where many have been laid off, made redundant or furloughed. 
\end{itemize}

Technology-related security risks are also a noteworthy concern. Exemplars are presented below.

\begin{itemize}
    \item \textit{Rushed technology adoption due to national lockdowns leading to the deployment of untested or unreliable technologies}. Such technologies may not work well and therefore give rise to employees adopting potentially dangerous shadow IT practices, e.g., not using Virtual Private Network (VPN) adequately, poor connection resulting in preferring insecure WiFi networks with better connection speeds, use of third-party services such as Dropbox, Google Drive for confidential work files.
    \item \textit{Unfamiliarity (or lack of proficiency) with new remote-working technology (e.g., Microsoft Teams, Zoom, etc.) leading to mistakes in the use and management of security features}. The speed at which these technologies have been implemented because of the pandemic places a technical burden on individuals, at a time when they are already in stressful and tense situations. 
    \item \textit{Security issues with remote-working and remote communication technologies can expose an organisation to increased risk}. As highlighted above, the rush in adoption of new platforms to operate during COVID-19 also exposed enterprises to a range of new threats accompanying such technologies. For instance, we have seen several attacks targeting Zoom and Microsoft over the last year.
    \item \textit{Intentional or inadvertent use of work devices for personal matters, therefore opening work devices to additional risk}. For instance, using work devices to watch films on illegitimate websites or download malicious attachments from personal emails, social media or gaming websites.
    \item \textit{Work devices may be stolen from the home or remote-working environment}. If these devices are not appropriately encrypted, they pose a risk to corporate data and services. This risk is increased during the pandemic because criminals are aware that most individuals are working remotely and therefore are likely to have more mobile technology at home.
    \item \textit{Employees returning to work after a long period of remote working may bring infected devices in to the corporate network}. Home networks are much more likely to be compromised than corporate networks, and therefore the extended period of remote working caused due to lockdowns, can increase the possibility of this risk. 
\end{itemize}

As can be seen from the examples above, security risks can originate from various areas. A primary difference with these risks in a post-COVID world is that they are exacerbated by the physical and mental impact that COVID-19 has had on people's lives.   

\section{Privacy Risks and Workplace Surveillance} 
\label{sec:PrivacyRisks}

While security risks and discussions dominated business concerns during the pandemic, privacy was a salient factor for employees. We analysed a series of current reports and articles exploring this issue, and noted an increasing prominence of discussions pertaining to workplace surveillance in remote-working setups. This was driven largely by employers and their worries about employee productivity, and secondly in an attempt to secure corporate data and systems. Below, we present exemplars of the primary risks to privacy emerging from our review. 

\begin{itemize}
    \item \textit{The potential infringement of employee's privacy caused by a dramatic surge in employer usage of (remote) workplace surveillance/monitoring technologies}. This could include monitoring of keystrokes, screens and websites visited (e.g., \cite{bloomberg2020rw}). A significant reality is that in some cases, employees may be using their own technologies (smartphones, iPads, laptops) for remote working, thereby giving employers---or the companies they outsource to---access to vast amounts of personal employees data.  
    \item \textit{New forms of technology emerging during the pandemic that are able to monitor employee emotional state (e.g., as smart technologies~\cite{bbcnews2020em}) could also violate privacy}. For example, such emotional and psychological data, if not properly protected, may be used to profile employees according to their well-being, and thus impact employment or future career prospects. 
    \item \textit{Exposure of personal information as a result of how remote working and communication technologies are used}. For instance, exposing home (living room, bedroom, office) backgrounds in video calls, or posting photos online of home offices can leak personal data (e.g., interest, hobbies)~\cite{threatpost2021ho} which can be further used as the basis for cybercrime. This touches on the common issue of oversharing online and its link to cyber risk~\cite{nursecyber-crime2019}; an issue overlooked in the pandemic as individuals focus primary on staying connected through online services.
\end{itemize}

\section{Discussion and Conclusion}
\label{sec:Conclusion}
A number of salient risks were identified in our reflection above, and it is important that organisations now reflect on the choices made during an emergency to keep the business going, and ensure they match their business as usual security posture and risk tolerance.  Where new software has been adopted there is now an opportunity to update training to remedy lack of proficiency issues, and review configurations to ensure any risks introduced during the pandemic are minimised. Most organisations have cyber security training for employees and it is important for this to include how to minimise security risks whilst working remotely. The training should cover taking sensible precautions to protect privacy, how to protect devices in public spaces (the home is not a public space but it may have untrusted people in it, so can, on occasion, be considered from this perspective), use of non-corporate networks, and the importance of using company sanctioned options for file transfer and access of company resources.  

Looking forward, another key risk is related to bringing infected devices back to the corporate network.  Prior to COVID-19 many organisations will have assumed that employee devices, which predominantly connect to the corporate network, are trusted devices and unlikely to be bringing infections into the network.  A separate guest network may also be available for external visitors.  It seems likely that in the future the default will be a Zero Trust architecture (where there is no trust in devices by default)~\cite{threatpost2020wfhs}. Therefore, so long as this remains usable for employees, this would eliminate the need to run multiple networks with different levels of trust, and would result in a greater level of security.  

Many organisations have indicated they do not expect employees to return to the office full time~\cite{bbc2020noplan}. There is now time to reflect on how to best support this way of working in the future.   To facilitate remote working, more employees will be issued with laptops, to avoid the need for them to use personal devices for work; though this does not address the issue of using work devices for personal activities. Whilst remote working is not new to most organisations, the move to more extensive home working also means employees need to be able to access a wide range of corporate systems and resources remotely in the long term. Now organisations should evaluate whether existing solutions meet future needs, and to identify alternatives if not. Once a solution has been chosen it is also vital to configure it correctly and ensure appropriate employee training is in place.  

Some risks are universal, including an increase in COVID-19-related phishing attacks and the need to keep software up to date, and everyone should be taking proactive steps to address them. Some of these steps may also include exploring the utility of cyber insurance, which can offer some level of protection and support after incidents~\cite{rusi2020ci}. Other risks, for example the likelihood of confidential conversations being overheard, are context specific, and this highlights the importance of asking people to consider the nature of their work and new work environment. Some jobs require a substantial number of video calls with external contacts, and for these workers the privacy related recommendations are more important than for someone only communicating with other employees. 

In terms of privacy, there is a clear tension between an employee’s wish for privacy in their own home, and the employer’s goal to be able to monitor productivity while workers are not co-located with them. Sometimes monitoring is introduced as a preventative measure, without there being any evidence to indicate it is needed.  There is limited evidence to show employees are doing less work whilst at home, in fact some research shows the opposite~\cite{bw2020prodr}, and the introduction of additional monitoring adds to the perception that employees are not trusted. A perception of a lack of trust coupled with intrusive monitoring may weaken the relationship between the organisation and such individuals.

The pandemic has increased acceptance of flexible work schedules, and this has been a substantial advancement in terms of inclusivity, particularly for those with caring responsibilities who might wish to complete their work according to their own schedule.  In order to maintain trust between employers and employees it is important to identify the minimum level of monitoring that could be used to give adequate assurance, and this will vary depending on the roles being undertaken. It is likely that cultural norms around levels of monitoring will develop, similar to expectations around realities such as time recording, which is commonplace in some sectors. 


\section*{Acknowledgements}
Funding for this research was received by SPRITE+: The Security, Privacy, Identity, and Trust Engagement NetworkPlus (EPSRC Grant reference EP/S035869/1).

\bibliographystyle{splncs04}
\bibliography{main}

\begin{thebibliography}{10}
\providecommand{\url}[1]{\texttt{#1}}
\providecommand{\urlprefix}{URL }
\providecommand{\doi}[1]{https://doi.org/#1}

\bibitem{bbcnews2020em}
{BBC}: {A wristband that tells your boss if you are unhappy} (2020),
  \url{https://www.bbc.co.uk/news/business-55637328} (Accessed on 14 March
  2021)

\bibitem{bbc2020noplan}
{BBC}: {No plan for a return to the office for millions of staff} (2020),
  \url{https://www.bbc.co.uk/news/business-53901310} (Accessed on 14 March
  2021)

\bibitem{bloomberg2020rw}
{Bloomberg}: {Bosses Panic-Buy Spy Software to Keep Tabs on Remote Workers}
  (2020),
  https://www.bloomberg.com/news/features/2020-03-27/bosses-panic-buy-spy-software-to-keep-tabs-on-remote-workers
  (Accessed on 14 March 2021)

\bibitem{buil2021cybercrime}
Buil-Gil, D., Mir{\'o}-Llinares, F., Moneva, A., Kemp, S.,
  D{\'\i}az-Casta{\~n}o, N.: Cybercrime and shifts in opportunities during
  covid-19: a preliminary analysis in the uk. European Societies
  \textbf{23}(sup1),  47--59 (2021)

\bibitem{bw2020prodr}
{Business Wire}: {Productivity Has Increased, Led By Remote Workers} (2020),
  \url{https://www.businesswire.com/news/home/20200519005295/en/}

\bibitem{green2020working}
Green, N., Tappin, D., Bentley, T.: Working from home before, during and after
  the covid-19 pandemic: implications for workers and organisations. New
  Zealand Journal of Employment Relations  \textbf{45}(2),  5--16 (2020)

\bibitem{ibm2020wfhs}
{IBM}: {IBM Security Work From Home Study} (2020),
  https://newsroom.ibm.com/2020-06-22-IBM-Security-Study-Finds-Employees-New-to-Working-from-Home-Pose-Security-Risk
  (Accessed on 14 March 2021)

\bibitem{infosecmag2020}
{Infosecurity Magazine}: {21\% of UK Workers Feel More Vulnerable to Cybercrime
  During COVID-19} (2020),
  https://www.infosecurity-magazine.com/news/uk-workers-vulnerable-cybercrime/
  (Accessed on 14 March 2021)

\bibitem{insider2019work}
{Insider}: {9 of the most challenging things about working remotely, according
  to people who do it} (2019),
  https://www.businessinsider.com/working-remote-challenges-work-from-home-2019-10
  (Accessed on 14 March 2021)

\bibitem{LALLIE2021102248}
Lallie, H.S., Shepherd, L.A., Nurse, J.R.C., Erola, A., Epiphaniou, G., Maple,
  C., Bellekens, X.: {Cyber Security in the Age of COVID-19: A Timeline and
  Analysis of Cyber-Crime and Cyber-Attacks during the Pandemic}. Computers \&
  Security (102248) (2021)

\bibitem{naidoo2020multi}
Naidoo, R.: A multi-level influence model of covid-19 themed cybercrime.
  European Journal of Information Systems  \textbf{29}(3),  306--321 (2020)

\bibitem{nursecyber-crime2019}
Nurse, J.R.C.: {Cybercrime and You: How Criminals Attack and the Human Factors
  That They Seek to Exploit}. In: The Oxford Handbook of Cyberpsychology (2019)

\bibitem{rusi2020ci}
Sullivan, J., Nurse, J.R.C.: {Cyber Security Incentives and the Role of Cyber
  Insurance (Royal United Services Institute (RUSI) Emerging Insights Paper)}
  (2020),
  https://rusi.org/publication/emerging-insights/cyber-security-incentives-and-role-cyber-insurance
  (Accessed on 14 March 2021)

\bibitem{threatpost2020wfhs}
{ThreatPost}: {Work-for-Home Shift: What We Learned} (2020),
  \url{https://threatpost.com/2020-work-for-home-shift-learned/162595/}

\bibitem{threatpost2021ho}
{ThreatPost}: {Home-Office Photos: A Ripe Cyberattack Vector} (2021),
  \url{https://threatpost.com/home-office-photos-cyberattack-vector/164460/}
  (Accessed on 14 March 2021)

\bibitem{tremblay2012telework}
Tremblay, D.G., Thomsin, L.: Telework and mobile working: analysis of its
  benefits and drawbacks. International Journal of Work Innovation
  \textbf{1}(1),  100--113 (2012)

\bibitem{waizenegger2020affordance}
Waizenegger, L., McKenna, B., Cai, W., Bendz, T.: An affordance perspective of
  team collaboration and enforced working from home during covid-19. European
  Journal of Information Systems  \textbf{29}(4),  429--442 (2020)

\bibitem{wilson2020job}
Wilson, J.M., Lee, J., Fitzgerald, H.N., Oosterhoff, B., Sevi, B., Shook, N.J.:
  Job insecurity and financial concern during the covid-19 pandemic are
  associated with worse mental health. Journal of occupational and
  environmental medicine  \textbf{62}(9),  686--691 (2020)

\bibitem{wef2020wfh}
{World Economic Forum}: {6 charts that show what employers and employees really
  think about remote working } (2020),
  https://www.weforum.org/agenda/2020/06/coronavirus-covid19-remote-working-office-employees-employers
  (Accessed on 14 March 2021)

\bibitem{whoimp2021}
{World Health Organisation (WHO)}: {Impact of COVID-19 on people's livelihoods,
  their health and our food systems} (2020),
  https://www.who.int/news/item/13-10-2020-impact-of-covid-19-on-people's-livelihoods-their-health-and-our-food-systems
   (Accessed on 14 March 2021)

\end{thebibliography}
\end{document}